\documentclass[prb,reprint,aps, superscriptaddress]{revtex4-2}
\usepackage[utf8]{inputenc}
\usepackage{hyperref}
\usepackage{natbib}
\usepackage{graphicx}
\usepackage{xcolor}

\graphicspath{{figures/}}
\usepackage[separate-uncertainty]{siunitx}
\usepackage{amsmath}
\renewcommand{\vec}[1]{\boldsymbol{#1}}
\begin{document}
%%%%%%%%%% Title %%%%%%%%%%
\title{Effective exchange interaction for terahertz spin waves in iron layers}
\date{\today}

%%%%%%%%%% Author/Affiliation %%%%%%%%%%
\author{L. Brandt}
\affiliation{Institute of Physics, Martin Luther University Halle-Wittenberg, 06120 Halle, Germany}
\author{U. Ritzmann}
\affiliation{Dahlem Center for Complex Quantum Systems and Fachbereich Physik, Freie Universit\"at Berlin, 14195 Berlin, Germany}
\author{N. Liebing}
\affiliation{Institute of Physics, Martin Luther University Halle-Wittenberg, 06120 Halle, Germany}
\author{M. Ribow}
\affiliation{Institute of Physics, Martin Luther University Halle-Wittenberg, 06120 Halle, Germany}
\author{I. Razdolski}
\affiliation{Faculty of Physics, University of Bialystok, 15-245 Bialystok, Poland}
\author{P. Brouwer}
\affiliation{Dahlem Center for Complex Quantum Systems and Fachbereich Physik, Freie Universit\"at Berlin, 14195 Berlin, Germany}
\author{A. Melnikov}
\affiliation{Institute of Physics, Martin Luther University Halle-Wittenberg, 06120 Halle, Germany}
\author{G. Woltersdorf} 
\affiliation{Institute of Physics, Martin Luther University Halle-Wittenberg, 06120 Halle, Germany}

%%%%%%%%%% Abstract %%%%%%%%%%
\begin{abstract}

    The exchange stiffness is a central material parameter of all ferromagnetic materials. Its value controls the Curie temperature as well as the dynamic properties of spin waves to a large extend. Using ultrashort spin current pulses we excite perpendicular standing spin waves (PSSW) in ultrathin epitaxial iron layers at frequencies of up to 2.4 THz. Our analysis shows that for the PSSWs the observed exchange stiffness of iron is about 20\% smaller compared to the established iron bulk value. In addition, we find an interface-related reduction of the effective exchange stiffness for layers with the thickness below 10 nm. To understand and discuss the possible mechanisms of the exchange stiffness reduction we develop an analytical 1D-model. In doing so we find that the interface induced reduction of the exchange stiffness is mode-dependent.
\end{abstract}

%%%%%%%%%% has to be after abstract %%%%%%%%%%
\maketitle

\section{Introduction}
\label{sec:introduction}
To date, neutron scattering measurements are considered to provide the best estimates of the spin-wave stiffness constant. The widely used value for bulk iron was established in 1968 by Shirane et al. \cite{Shirane1968} to be D = \SI{281+-10}{\milli \eV \square \angstrom } using a macroscopic Fe single crystal. Again, using neutron scattering, this result was confirmed by Mook et al. \cite{Mook1973}. The advantage of neutron scattering experiments is the ability to probe very large wave vectors (up to the Brillouin zone boundary) compared to most other methods. By performing the experiment along different crystal orientations, the spin wave dispersion can be determined for all directions. Unfortunately, large single crystals of bcc iron required for neutron scattering (inch sized) are difficult to prepare with high quality. Furthermore, one should point out, that the stiffness constants were obtained using Taylor expansion for the spin wave dispersion up to $k^2$ or $k^4$. This, and the presence of Stoner excitations overlapping with the spin wave band can cause substantial uncertainties at large wave vectors. Hicken et al. \cite{Hicken1995} reported a significantly reduced stiffness constant for Fe layers with decreasing film thickness. They studied perpendicular standing spin-waves (PSSW)  by means of Brillouin light scattering (BLS) in bcc Fe films for a thickness range of 2-117 nm epitaxially grown on GaAs(001). The stiffness constant of the thickest sample was found to be D = \SI{260}{\milli \eV \square \angstrom}  and hence close to the bulk value determined from neutron scattering measurements. As the samples of this study also showed some degree of chemical interdiffusion at the Fe/GaAs interface, the authors concluded that the stiffness may be related to the interfaces.  Razdolski et al.  \cite{Razdolski2017} also found a reduced stiffness constant of D $\approx$ \SI{200}{\milli \eV \square \angstrom } for a 12.7 nm epitaxial grown Fe film. In this case, the Fe thickness and the sharpness of interfaces were confirmed by cross sectional transmission electron microscopy. Furthermore, using spin polarised electron energy loss spectroscopy (SPEELS) also reduced stiffness of only \SI{160}{\milli \eV \square \angstrom } was observed for a 3 nm thick Fe film \cite{Prokop2009}.  More recently, Lalieu et al. \cite{Lalieu2017} investigated optically excited PSSWs in thin Co films. Here an exponentially decreasing exchange stiffness constant with decreasing the thickness from 6 nm to 3 nm was found and attributed to chemically intermixed interface regions in the sputtered Co samples. It is worthwhile to point out, that in this work the spin-current is excited in the same magnetic layer whose dynamics is studied.  In this case an interplay of demagnetization and spin wave dynamics can be expected. In a follow up work by the same group  Lalieu et al. investigated optically excited PSSWs in the THz frequency range and observed no thickness dependence of the exchange stiffness parameter\cite{Lalieu2019}.

Besides the experimental work there has also been a substantial theoretical effort to determine the exchange stiffness with ab-initio methods. Unfortunately, the numerical results of these calculations vary significantly (particularly for the case of Fe). Pajda et al. showed  \cite{Pajda2001} that origin for this behavior is most likely caused by the fact that the exchange interaction has a long range oscillatory character in Fe (in contrast to Ni or Co) and therefore it is not sufficient to consider only the next nearest neighbor -interactions. Pajda et al. also provided a regularization procedure to estimate the stiffness constant and got the value \SI{250+-7}{\milli \eV \square \angstrom} for Fe case. The work of Sipr et al. \cite{Sipr2020} summarizes the recent efforts for ab-initio calculations of the spin wave stiffness and illustrates the crucial influence of various technical parameters. The resulting calculated spin wave stiffness constants for bulk iron at 0 K vary between 262 and \SI{302}{\milli \eV \square \angstrom }.  Regarding the reduction of effective stiffness for ultrathin films, the authors are not aware of any microscopic description. Therefore the development of such models is highly desirable. 

In the present work we approach the exchange stiffness parameter with a new experimental method providing access to the dynamics of spin-waves with wave length in the nanometer range. The corresponding wave vectors are located between the values relevant for typical GHz magnetization dynamics experiments such as BLS (a few hundred nm wavelengths) and the large wave vectors usually only accessible by neutron scattering experiments (1~nm wavelength) as illustrate in Fig.~\ref{fig:sample}(a). This intermediate wave-vector range has the advantage that higher order terms of the dispersion as well as Stoner excitations can be neglected in the spin wave dispersion. For this, we measure in the ultrafast time-domain the magnetization in Fe layers excited by ultrashort spin current pulses. Our results consistently point towards a reduced stiffness constant of only \SI{220}{\milli \eV \square \angstrom } in bulk iron. An additional reduction mechanism occurs for samples with a thickness below 8~nm. The interface induced reduction of the stiffness is modelled by interface disorder on the atomic scale using an analytical model as well as atomistic simulations of the spin dynamics.

The manuscript is organized as follows: First we will introduce our experimental method in Sec. \ref{sec:experiment} and summarize the results in Sec. \ref{sec:results}. Then a theoretical model capable of describing the lowered stiffness of the thinnest samples is introduced in Sec. \ref{sec:modelling} . Finally, the experiments are interpreted using the theoretical description in Sec. \ref{sec:modelresults}.

\section{Experimental approach}
\label{sec:experiment}
\begin{figure}[htbp]
	\centering
	\begin{minipage}[]{0.42\textwidth}	
		\raggedright (a)
		\includegraphics[width=1\linewidth]{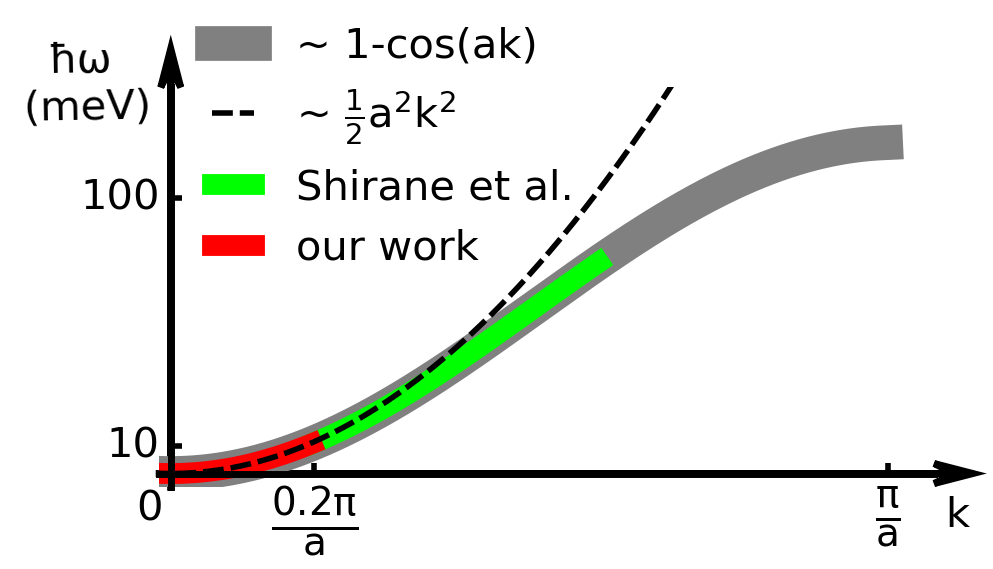}
    \end{minipage}
    
    \begin{minipage}[]{0.42\textwidth}	
        \bigskip
	    \raggedright (b)
		\includegraphics[width=1\linewidth]{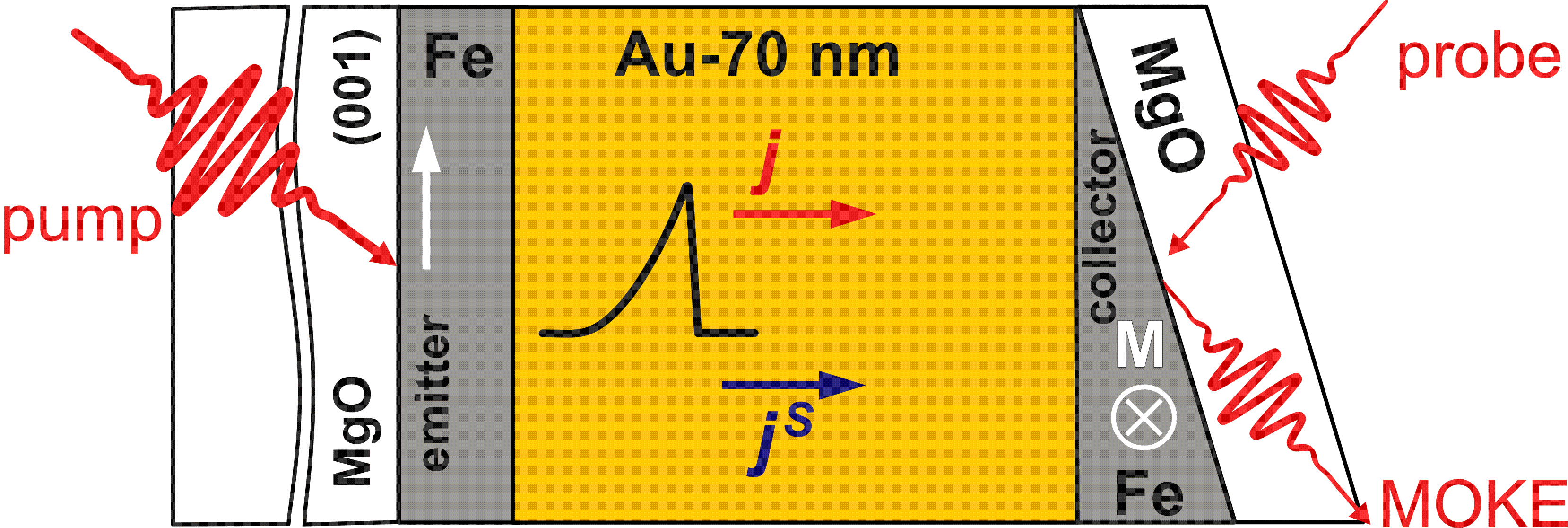}
    \end{minipage}
	\caption{ (a) Schematic spin wave dispersion of Fe in the Brillouin zone with indicated measurement ranges for neutron scattering (e.g. \cite{Shirane1968}) and our work. (b) Sample layout and optical pump probe measurement geometry.}
	\label{fig:sample}
\end{figure}

The experimental configuration is illustrated in Fig.~\ref{fig:sample}(b). A femtosecond laser pulse excites the electrons in ultrathin Fe layer and launches an intense spin current pulse into an adjacent Au layer. The spin current is absorbed by second ferromagnetic layer and exerts an ultrashort and intense spin torque pulse, capable of exciting magnetization dynamics in the THz frequency range. The experimental setup and the measurement technique are described in detail in  \cite{Alekhin2017,Razdolski2017}. A back pump-front probe approach is used \cite{Melnikov2011, Alekhin2017,Razdolski2017}. The sample we study here has a thickness gradient of the second Fe layer allowing systematic measurements of frequencies of PSSW modes vs.~ the thickness of the Fe layer.

The sample (Fig.~\ref{fig:sample}(b)) has been grown by means of molecular beam epitaxy in ultrahigh vacuum at a base pressure \SI{3e-11}{\milli\bar}. A double-side polished MgO(001) substrate was annealed at \SI{500}{\degreeCelsius}. Prior to the metal layer deposition a \SI{10}{\nano\metre} thick epitaxial MgO buffer layer was deposited. On this substrate, a \SI{4.4}{\nano\metre}-thick Fe(001) layer has been grown, which serves as spin current emitter in the  experiments. Then, the \SI{70}{\nano\metre} Au(001) layer was deposited. On top of the Au spacer layer, the second Fe(001) layer (collector) was deposited, as a wedge with the thickness between \SIrange{1}{17}{\nano\metre}. Finally, the sample has been capped with \SI{10}{\nano\metre} MgO(001) and  \SI{25}{\nano\metre} Al$_2$O$_3$ for protection at ambient conditions.
During the growth of MgO buffer and Fe emitter layers reflection high-energy electron diffraction (RHEED) oscillations were observed, demonstrating a good layer-by-layer growth. 

In the time resolved optical experiments a cavity-dumped Ti:sapphire oscillator (Mantis, Coherent) is used. This laser operates at 800 nm and has a pulse length of 14 fs with a repetition rate of 1 MHz. The light was split at a power ratio 4:1 into pump and probe beams. The measurements were performed at room temperature. A crossed pair of Helmholtz coils provides magnetic fields up to 5 mT in the sample plane.

For the optical experiments it is essential that the two magnetic layers of the layer stacks shown in Fig.~\ref{fig:sample}(b) can be switched independently at different magnetic fields \cite{Razdolski2017,Alekhin2017}. This is necessary in order to suppress unwanted signal contributions in the time traces. The idea is as follows: the direction and thereby the phase of magnetization precession in the collector is determined by the direction of magnetization in collector and by the orientation of the injected spins. The latter is determined by the emitter magnetization. Therefore, in the presence and absence of the pump beam the magneto-optical Kerr effect (MOKE) signals for every step of the pump-probe delay scan are obtained for four magnetic states with emitter and collector magnetizations directed up and right (UR), down and right (DR), up and left (UL), and down and left (DL), respectively. Following \cite{Razdolski2017} and adding the pump-induced variations of these signals as (UR-DR-UL+DL)/4 we obtain the noise-reduced and background-free oscillatory dynamics of the polar MOKE rotation and ellipticity signals. See more details in Appendix \ref{sec:appa}.

\section{Experimental results}
\label{sec:results}
\begin{figure}[htbp]
	\centering
	\begin{minipage}[]{0.42\textwidth}	
		\raggedright	(a)
		\includegraphics[width=1\linewidth]{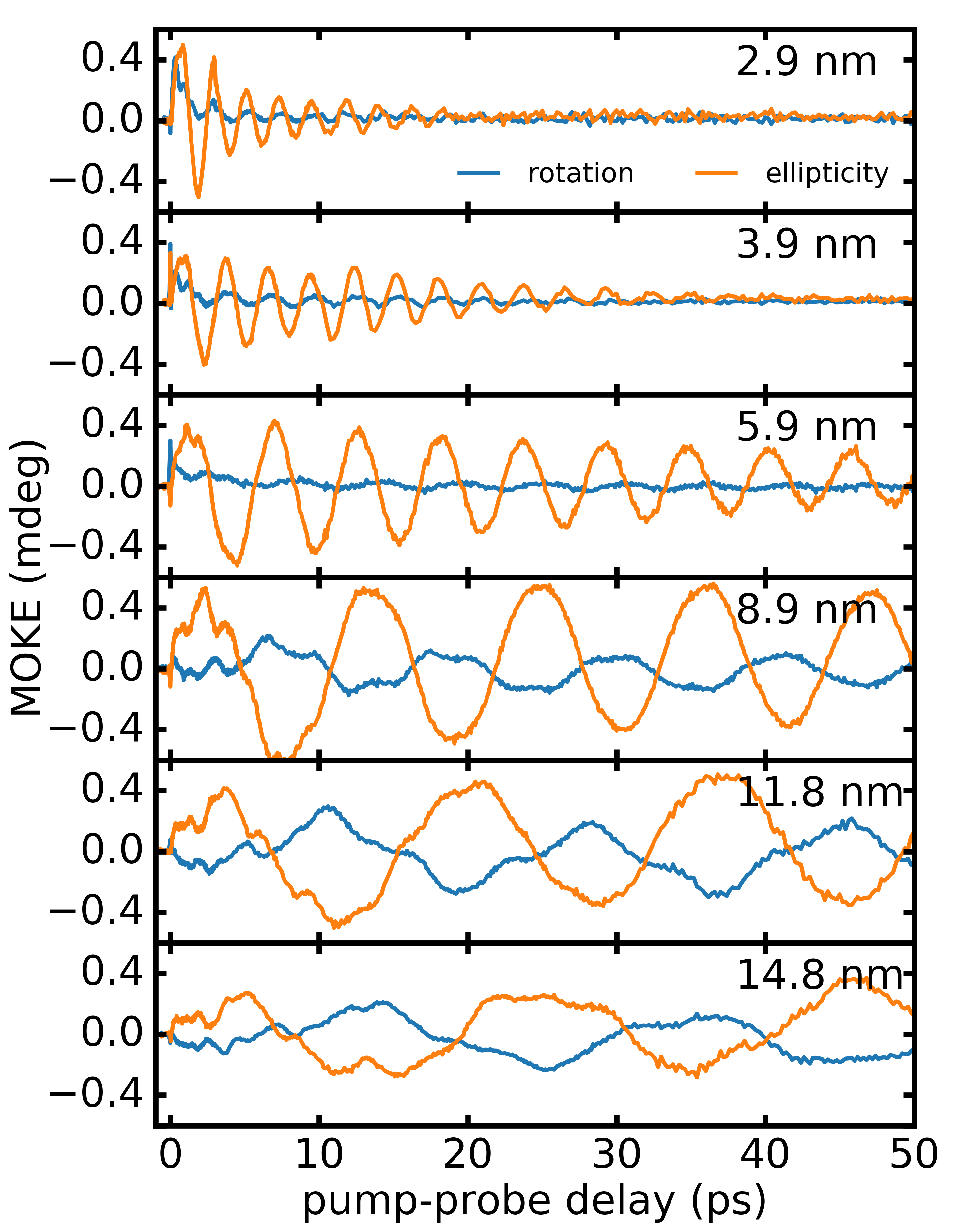}
    \end{minipage}
    \begin{minipage}[]{0.42\textwidth}	
        \raggedright	(b)
		\includegraphics[width=1\linewidth]{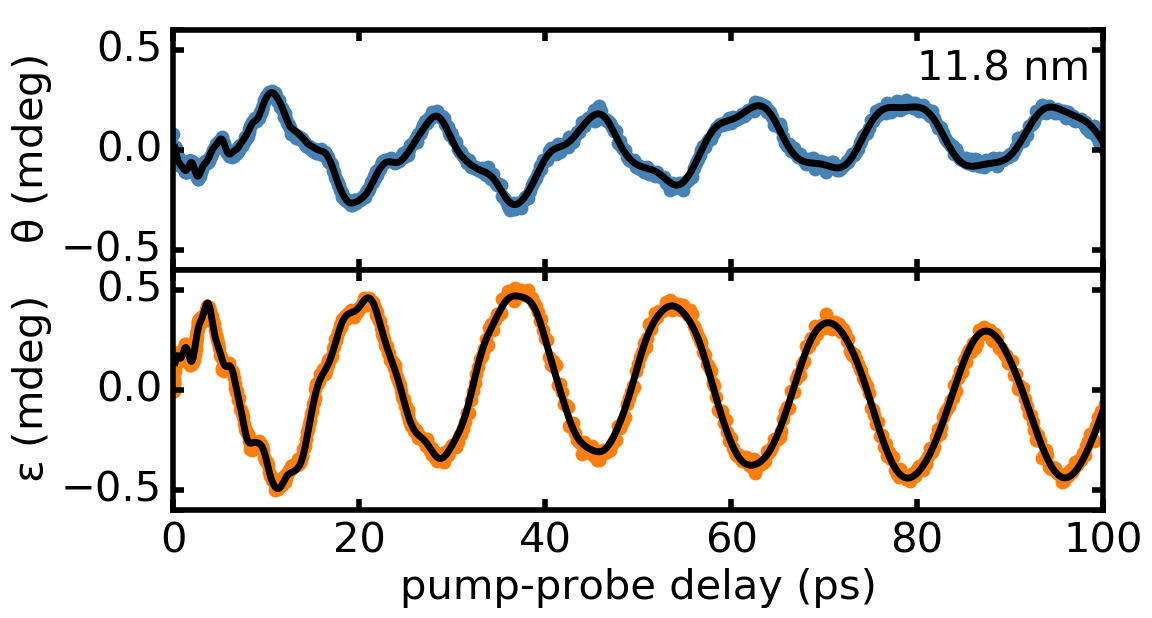}
    \end{minipage}   
    \begin{minipage}[]{0.42\textwidth}	
        \raggedright	(c)
		\includegraphics[width=1\linewidth]{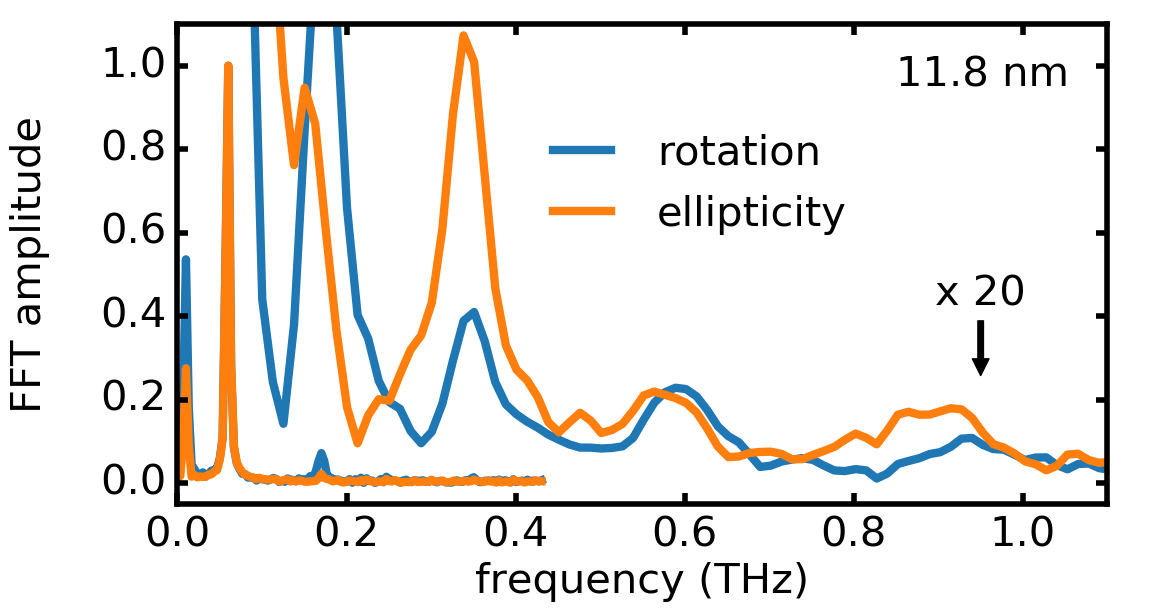}
    \end{minipage}
	\caption{(a) Time-resolved polar MOKE rotation and ellipticity data measured at several thicknesses of the Fe collector. (b) Sum of damped cosines fit (black line) to MOKE rotation and ellipticity data exemplified for 11.8 nm-thick collector; (c) FFT amplitude (normalized to the amplitude of first PSSW mode) of MOKE rotation and ellipticity data shown in (b). FMR mode as well as 2 PSSW modes are visible for the long time range measurement ($t_{max}$ = 700 ps, $\Delta t$ = 0.2 ps) and another 3 PSSW modes are visible for the short time range measurement ($t_{max}$ = 30 ps, $\Delta t$ = 0.02 ps)}
	\label{fig:raw}
\end{figure}

In Fig. \ref{fig:raw}(a) the measured time-resolved polar MOKE signals (rotation $\theta$ and ellipticity $\varepsilon$) are shown. It should be noted, that the amplitudes of the measured oscillations strongly depend on the depth profile of the polar MOKE sensitivity \cite{Razdolski2017}. The time trace of the damped periodic signals contain multiple frequency components indicating the excitation of several perpendicular standing spin-wave eigenmodes. These distinct modes are clearly visible in the frequency domain obtained by fast Fourier transformation (FFT) of the time domain data as shown in Fig. \ref{fig:raw}(c). By fitting the time domain data (Fig. \ref{fig:raw}(b)) with a sum of damped cosine functions

\begin{equation}
\theta,\varepsilon = \sum_n A^{\theta,\varepsilon}_n \cos{[2\pi f_n t + \varphi_n]} e^{-\frac{t}{\tau_n}}
\label{eq:cos}
\end{equation}

one obtains a set of amplitudes $A^{\theta,\epsilon}_n$, phases $\varphi_n$, lifetimes $\tau_n$ and frequencies $f_n$ for at least 6 modes (FMR mode and 5 PSSW modes) as a function of the collector thickness $d_{\mathrm{Fe}}$.

Neglecting the dipolar interaction, the spin-wave frequency $f$ for a thin magnetic film can be described as follows \cite{Gurevich1996}:
\begin{equation}
f(k) = \gamma \sqrt{(\tilde{H}_{\mathrm{an}} + \tilde{D} k^2) (\tilde{H}_{\mathrm{an}} + \tilde{H}_{\mathrm{dem}} + \tilde{D} k^2)}.
\label{eq:spinwave}
\end{equation}
For PSSWs, the wave vectors are quantized and given by
\begin{equation}
k=\frac{n\pi}{d_{\mathrm{Fe}}} \; ,
\label{eq:k}
\end{equation}
where the PSSW mode number $n$ is an integer representing the number of nodes. Here the widely accepted approximation of free boundary conditions is used owing to the non-magnetic adjacent layers and the absence of magnetization pinning. The spin-wave stiffness is represented by $\tilde{D}$. Due to the combination of bulk and interface properties in ultrathin magnetic films the thickness dependence of anisotropy and demagnetization fields is described as follows \cite{Hallal2013,Yang2011}: 

\begin{equation}
\tilde{H}_{\mathrm{an}} = H_{\mathrm{an}} \left(1- \frac{d_{\mathrm{an}}^0}{d_{\mathrm{Fe}}}\right), \tilde{H}_{\mathrm{dem}} = H_{\mathrm{dem}} \left(1- \frac{d_{\mathrm{dem}}^0}{d_{\mathrm{Fe}}}\right)
\label{eq:Handem}
\end{equation}

\begin{figure}[htbp]
	\centering
	\includegraphics[width=0.9\linewidth]{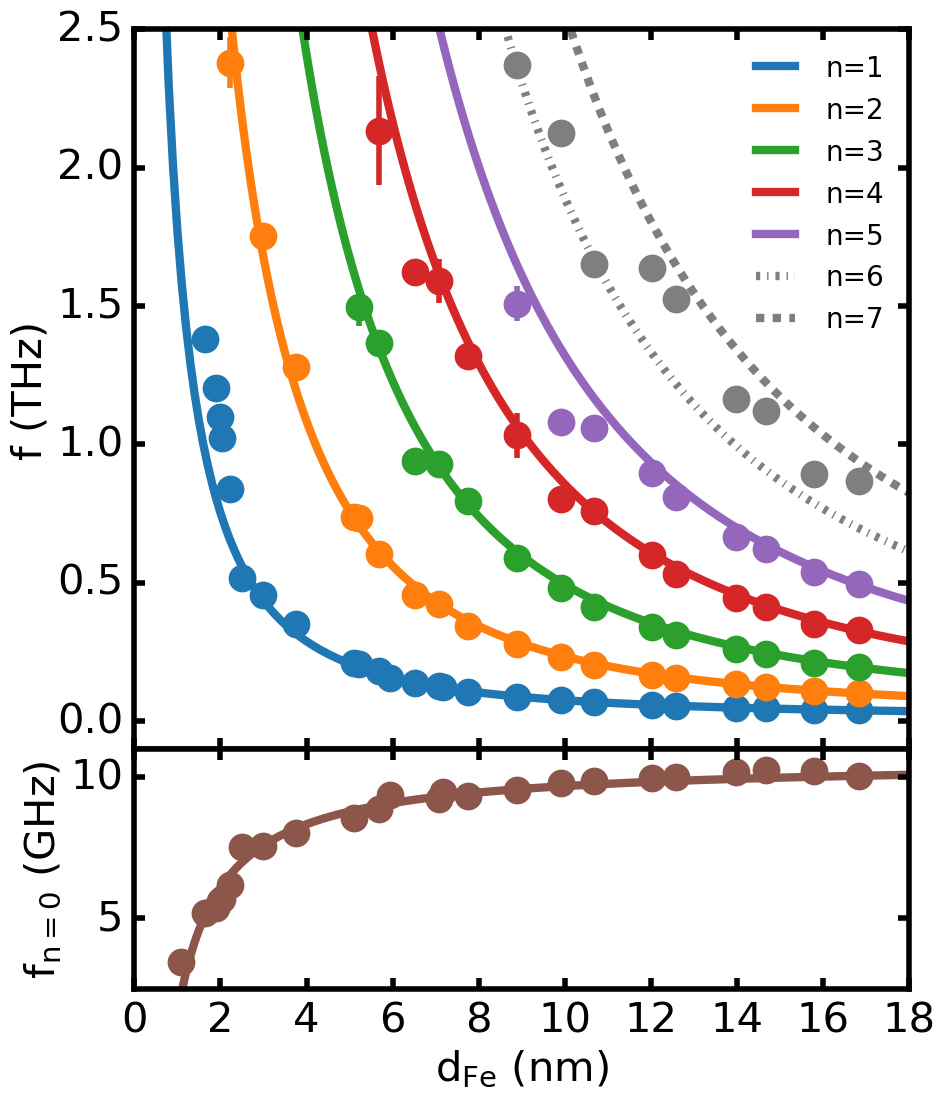}
	\caption{(lower graph) FMR and (upper graph) PSSW frequencies  obtained by  fits to the tr-MOKE traces (Eq. (\ref{eq:cos})). The error bars are given by the fit procedure. The solid lines are a fit to the Kittel-formula (Eq. (\ref{eq:spinwave})) using an exponential thickness dependence of the stiffness (Eq. (\ref{eq:expD})) with $D^{\infty}=220 \, \mathrm{meV\mathring{A}^2}$. Due to their large error bars in the frequencies of the 6th and 7th mode these modes were not included in the fit.}
	\label{fig:spinwave}
\end{figure}

The values of $\mu_0 H_{\mathrm{dem}}$ = 2.17 T, $\mu_0 H_{\mathrm{an}}$ = 0.064 T, $d_{\mathrm{an}}^0$ $\approx$ $d_{\mathrm{dem}}^0$ = 0.85 nm are determined by static MOKE measurements (see Appendix \ref{sec:appb}) and consistent with the fit of the ferromagnetic resonance frequency as a function of thickness (see Fig.~\ref{fig:spinwave} lower part).

\begin{figure}[htbp]
	\includegraphics[width=0.9\linewidth]{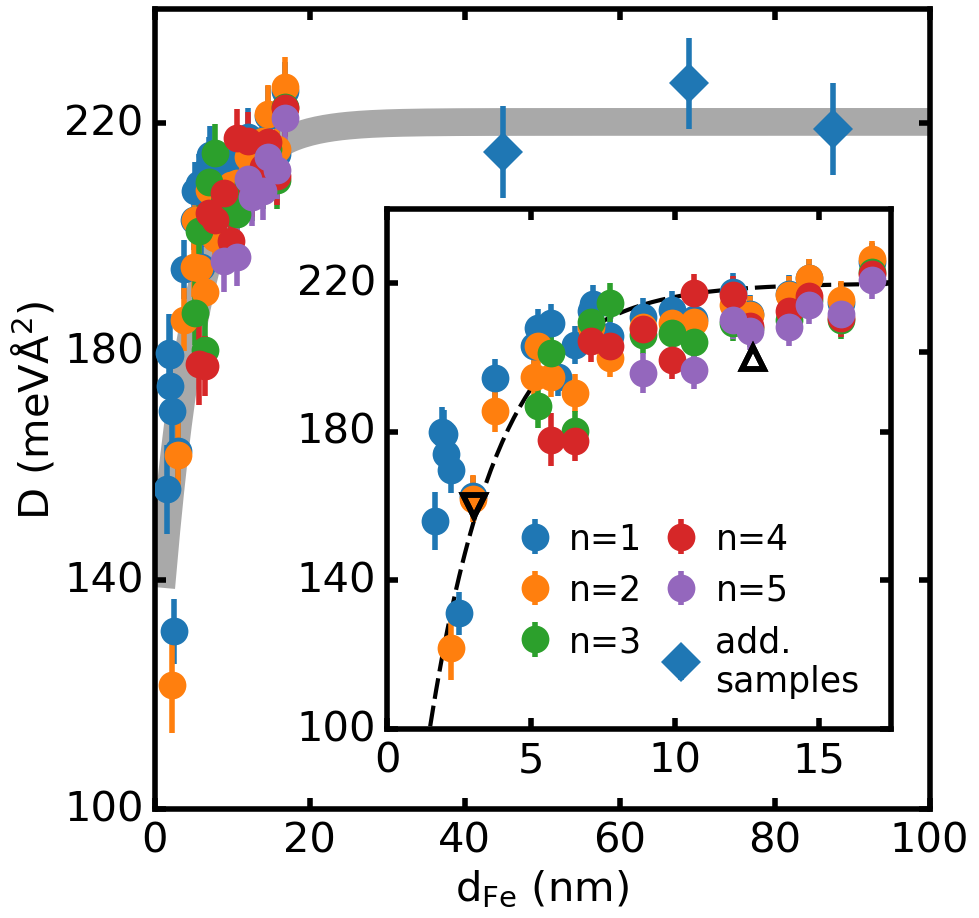}
	\caption{Effective stiffness of PSSWs measured by tr-MOKE (points). The diamonds are the PSSW results obtained with the three additional samples using FMR (only first mode). The dashed line in the inset represents  the exponential fit according to Eq. (\ref{eq:expD}) corresponding to a bulk stiffness of $D^{\infty} = 220$ $\mathrm{meV\mathring{A}^2}$ and a critical thickness of $d_{D}^0$ = 2.4 nm resulting from fit by Kittel-formula (Eq. (\ref{eq:spinwave})). The triangles show the literature values obtained by Prokop et al. \cite{Prokop2009} (triangle down) and Razdolski et al. \cite{Razdolski2017} (triangle up). The gray thick line are a guide to the eye.}
	\label{fig:stiff}
\end{figure}

By extracting the effective stiffness values as a function of Fe thickness from the data shown in Fig.~\ref{fig:spinwave} we find that it only reaches a value of about $220 \, \mathrm{meV\mathring{A}^2}$ (Fig.~\ref{fig:stiff}).
In order to confirm that the exchange stiffness for the 15 nm iron film in the optical experiment indeed already converges towards the bulk value we performed additional experiments with even thicker Fe layers. For this, three samples with  45 nm, 69 nm, 87.5 nm thick Fe single layers were deposited by means of magnetron sputtering onto annealed MgO(001) substrates. RHEED patterns confirmed that these Fe layers are single crystalline and their roughness is comparable to that of the MBE grown samples. In-situ capping by 3 nm Au protected these samples from oxidation. The iron thickness of the three samples was determined accurately by Rutherford backscattering spectrometry (RBS). Using broadband ferromagnetic resonance \cite{Woltersdorf2009,Obstbaum2014} the homogeneous mode and first PSSW mode were measured between 2 and 26 GHz with the magnetic field applied in-plane. Fits to the frequency-dependent resonance positions of FMR and first PSSW modes allowed to determine the spin stiffness and the results are also indicated in Fig. \ref{fig:stiff} as blue diamonds. See more details of resonance measurements in Appendix \ref{sec:appc}. These independent results confirm that the spin stiffness in iron indeed converges towards a bulk value of only $220 \, \mathrm{meV\mathring{A}^2}$, which is about 20\% smaller than the literature values.

Besides the low bulk stiffness value for thinner Fe layers with $d_{\mathrm{Fe}} \lesssim $ 10 nm an additional thickness dependence is observed for the spin stiffness $D(d_{\mathrm{Fe}})$ (Fig. \ref{fig:stiff}).
Lalieu et al. \cite{Lalieu2017} (sputtered Co layers) and Hicken et al. \cite{Hicken1995} (evaporated Fe layers) explained similar observations of reduced exchange stiffness with decreasing FM thickness by chemical intermixing at the interfaces.

In order to analyze the thickness dependence of spin-wave stiffness from the PSSW data one can follow the phenomenological approach by Lalieu et al.~\cite{Lalieu2017} to treat a decreasing spin stiffness with decreasing thickness due to the influence of the interfaces:
\begin{equation}
\tilde{D} = D^{\infty} \left(1- \mathrm{exp}\left[-\frac{d_{\mathrm{Fe}}}{d_{D}^0}\right]\right).
\label{eq:expD}
\end{equation}
As we have already determined the bulk stiffness to be  $220 \, \mathrm{meV\mathring{A}^2}$ the only remaining fit parameter is the critical thickness $d_{D}^0$. This parameter characterizes the thickness when the stiffness reduced to a value of 1/e by interface effects such as intermixing or modified exchange coupling. 

As one can see in Fig.~\ref{fig:spinwave} the spin wave frequencies from the time-resolved MOKE measurements are in general well described by the theoretical dispersion relation (Eq. (\ref{eq:spinwave}) resulting in a value of 2.4 nm for the the critical thickness. However, a deviation is clearly observed at smallest used thicknesses for the first PSSW mode (blue points). Obviously the phenomenological description is not sufficient for ultrathin layers as highlighted in the inset of Fig. \ref{fig:stiff} where the fit result is reproduced by the dashed line. In addition one should keep in mind that $d_{D}^0$ is a purely phenomenological concept and cannot be directly related to the interface structure. For this a microscopic model is required.

In the present case the samples were grown by means of molecular beam epitaxy on MgO substrates. In this case very sharp interfaces with a minimal degree of intermixing are expected. Transmission electron microscopy studies showed in similar samples \cite{Alekhin2017,Razdolski2017,Melnikov2011} the interfaces between Fe and Au or Fe and MgO are nearly perfect. Therefore these sample represent a promising case to develop and verify a microscopic model. 
To explain the lowered bulk value of exchange stiffness, one would need to refine the calculation of the absolute values of the exchange integrals. 
In particular for Fe this is a long standing theoretical problem in terms of modelling and beyond the scope of this manuscript. Here we will focus on the thickness dependence of the stiffness and explore its connection to the atomic interface intermixing/disorder.

\section{Modelling of the spin wave stiffness in thin layers}
\label{sec:modelling}
A microscopic understanding of the interface-induced reduction of the spin stiffness is highly desirable. For this we develop the following model.
Consider magnetic iron with a bcc lattice with normalized magnetic moments of magnitude $S_{\vec{r}} = 1$ and direction $\vec{e}_{\vec{r}}$ on lattice sites $\vec{r} = (x, y, z)$. The magnetic moments are subject to a Heisenberg exchange interaction with exchange constant $J_{rr'}= J_{r'r}$ up to the 8th nearest neighbor. In addition, we consider an easy-axis anisotropy energy $d_x$ in $x$-direction and a hard-axis $d_z$ in $z$-direction (out-of-plane). The exchange parameters and anisotropy constants we use are shown in table \ref{tab:input}.

\squeezetable
\begin{table}
	\centering
    \begin{tabular}{c S[table-format=1.3]}
	\hline\hline
	parameter &  {value }  \\ 
	\hline
    $J_1$ (mRy) & 1.24\cite{Mryasov1996}   \\ 
    $J_2$ (mRy) & 0.646\cite{Mryasov1996}   \\ 
    $J_3$ (mRy) & 0.007\cite{Mryasov1996}   \\ 
    $J_4$ (mRy) & -0.108\cite{Mryasov1996}  \\
    $J_5$ (mRy) & -0.071\cite{Mryasov1996}   \\
    $J_6$ (mRy) & 0.035\cite{Mryasov1996}   \\
    $J_7$ (mRy) & 0.002\cite{Mryasov1996}   \\
    $J_8$ (mRy) & 0.014\cite{Mryasov1996}   \\
    $d_x$ ($\mathrm{\mu}$eV) & 6.97\cite{Ritzmann2020,Razdolski2017}   \\
    $d_z$ (meV) & -0.267\cite{Ritzmann2020,Razdolski2017}   \\
    \hline\hline
	\end{tabular}
	\caption{Table summarizing the input parameters for analytical model and atomistic spin dynamics simulations. $J_n$ denote exchange integrals for the interaction with ions at the $n$th coordination sphere.}
	\label{tab:input}
\end{table}

\begin{figure}[htbp]
	\centering
	\begin{minipage}[]{0.42\textwidth}	
		\raggedright	(a)
		\includegraphics[width=1\linewidth]{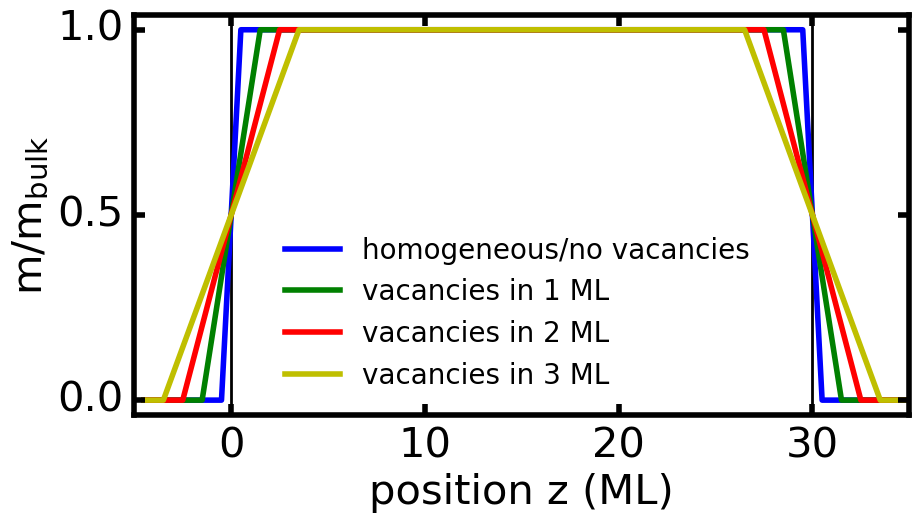}
	\end{minipage}	
	\begin{minipage}[]{0.42\textwidth}	
		\raggedright	(b)
		\includegraphics[width=1\linewidth]{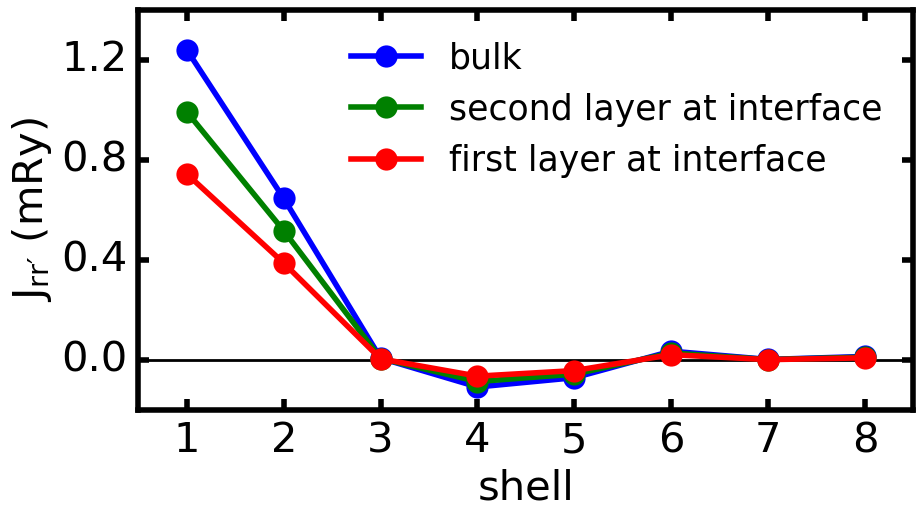}
	\end{minipage}	
	\caption{(a) Normalized magnetization as a function of thickness for a magnet without vacancies and three cases with a redistribution of magnetic moments in  1 ML, 2 ML and 3 ML around the boundary. (b) Exchange parameters as a function  of shell number, reduced to 80 \% for second layer and to 60 \% for first layer to interface, respectively.}
	\label{fig:U1}
\end{figure}

In the following, we will develop an effective one-dimensional model to describe magnon eigenmodes in thin magnetic films. We study the impact of modifications of the magnet at the boundary including vacancies and variation of the exchange constant. In Fig. \ref{fig:U1}(a) the magnetization is shown as a function of the position across the layer thickness for the case of no vacancies and for two cases with redistribution of the magnetic moment from the first monolayer or from the the first two monolayers around the film boundary, respectively. The magnetization increases in the latter two cases linearly with a constant slope around the interface. Note that the distance between monolayers (ML) corresponds to half of the lattice constant. To implement the spatial dependence of the magnetization, we consider a bcc lattice with randomly selected vacancies close to the magnetic layer boundary. We determine the thickness-dependence of the magnon frequencies and the effective exchange constant in these systems. Additionally, we will perform atomistic spin dynamics simulations of the magnetic films with ultrafast laser-induced spin-transfer torques exciting the magnetic thin film. As a last step, we explore the impact of interface modifications of the exchange interaction parameters on the magnon modes and the resulting effective exchange stiffness.

%\subsection{Effective one-dimensional model}
As a first step, we will derive an analytical model to describe magnon eigenmodes of the given magnetic system. In the absence of damping, the equation of motion is given by
\begin{equation}
\frac{d\vec{S_{\vec{r}}}}{dt} = - \frac{1}{\hbar} \vec{S}_{\vec{r}} \times \left( \sum_{\vec{r}'} J_{\vec{r}\vec{r}'} \vec{S}_{\vec{r}'} + 2 d_x S_{\vec{r}}^x \hat{\vec{x}} \right)   
\label{eq:U1}
\end{equation}
where we consider only the easy-axis anisotropy and set $d_z = 0$. Note that this assumption simplifies the following calculations, but it has no effect on the calculated exchange stiffness in the magnet. We consider a thin film with thickness d in z-direction and periodic boundary conditions in x- and y-directions. We look for periodic solutions of the equation of motion with the ansatz
\begin{equation}
\vec{S}_{\vec{r}} (t) = S_{\vec{r}} \vec{e}_z(t)
\label{eq:U2}
\end{equation}
i.e., all spins with the same $z$ coordinate have the same direction $\vec{e}_z$. By inserting the ansatz \ref{eq:U2} into the equation of motion \ref{eq:U1} we can construct an effective one-dimensional model. Hereto we proceed by constructing the equation of motion for the mean spin $\bar{\vec{S}}_z$ averaged over all lattice sites with the same $z$ coordinate,
\begin{equation}
\bar{\vec{S}}_z = N^{-1} \sum_{x,y} \vec{S}_{\vec{r}}
\label{eq:U3}
\end{equation}
where N is the number of lattice sites in a cross section
at fixed z,
\begin{equation}
\begin{split}
\frac{d\bar{\vec{S}}_z}{dt} &= \frac{1}{N} \sum_{x,y} \frac{d\vec{S}_{\vec{r}}}{dt}\\\  &= - \frac{1}{\hbar} \bar{\vec{S}}_z \times \left( \sum_{z'} \bar{J}_{zz'} \bar{\vec{S}}_{z'} + 2 d_x \bar{S}_{\vec{r}}^x \hat{x} \right)
\end{split}
\label{eq:U4}
\end{equation}
where
\begin{equation}
\bar{J}_{zz'} = \frac{\sum_{x,y} \sum_{x',y'} J_{\vec{r}\vec{r}'} S_{\vec{r}} S_{\vec{r}'} } {N \bar{S}_z \bar{S}_{z'} }
\label{eq:U5}
\end{equation}
Note that $\bar{J}_{zz'}$ only contains contributions where $z \neq z'$. The other contributions are not relevant as driving effective field in Eq. (\ref{eq:U4}). This is the equation of motion for a one-dimensional model with spins $\bar{S}_z = \bar{S}_z e_z$ and exchange interaction $\bar{J}_{zz'}$. The Hamiltonian for this system is
\begin{equation}
\mathcal{H} = - \sum_{z,z'} \bar{J}_{zz'} \bar{\vec{S}}_z \cdot \bar{\vec{S}}_{z'} - \sum_z d_x ( \bar{\vec{S}}_z \cdot \hat{x} )^2
\label{eq:U6}
\end{equation}
As a next step, we perform a Holstein-Primakoff transformation and parameterize
\begin{equation}
\bar{\vec{S}}_z = \vec{e} \sqrt{\bar{S}_z^2 - \bar{S}_z \bar{s}_z^2} + \bar{\vec{s}}_z \sqrt{\bar{S}_z}
\label{eq:U7}
\end{equation}
where $\bar{S}_z$ is the averaged magnitude of the magnetic moments at lattice position $z$ and $\bar{\vec{s}}_z$ the normalized transverse magnetization amplitude. The vector $\vec{e}$ points along the equilibrium magnetization direction. Expanding $\mathcal{H}$ to quadratic order in the amplitudes $\bar{\vec{s}}_z$ results in the magnon Hamiltonian $\mathcal{H}^{\mathrm{mag}}$,
\begin{equation}
\mathcal{H}^{\mathrm{mag}} = \frac{1}{2} \sum_{z,z'} J_{z,z'} | \bar{\vec{s}}_z \sqrt{\bar{S}_{z'}} - \bar{\vec{s}}_{z'} \sqrt{\bar{S}_z} |^2 + \frac{1}{2} d_x \sum_{\vec{r}} \vec{s}_{\vec{r}}^2
\label{eq:U8}
\end{equation}
The normal modes can be identified by writing $\mathcal{H}^{\mathrm{mag}}$ in a quadratic form,
\begin{equation}
\mathcal{H}^{\mathrm{mag}} = \frac{1}{2} \sum_{z,z'} \bar{\vec{s}}_z \cdot H_{zz'} \bar{\vec{s}}_{z'}
\label{eq:U9}
\end{equation}
where the matrix $H_{zz'}$ is real and symmetric. Upon shifting to the basis of normalized eigenvectors $v_{\mu,\vec{r}}$ of $H$, which are labeled by the index $\mu$, one has
\begin{equation}
\mathcal{H}^{\mathrm{mag}} = \frac{1}{2} \sum_{\mu} \omega_{\mu} |\vec{s}_{\mu}|^2
\label{eq:U10}
\end{equation}
where
\begin{equation}
\vec{s}_{\mu} = \sum_{\vec{r}} v_{\mu,\vec{r}} \vec{s}_{\vec{r}}
\label{eq:U11}
\end{equation}
The equation of motion
\begin{equation}
\begin{split}
\frac{d\vec{s}_{\mu}}{dt} 
&= - \frac{1}{\hbar} \vec{e} \times \frac{\partial \mathcal{H}^{\mathrm{mag}}}{\partial \vec{s}_{\mu}}\\
&= - \omega_{\mu} \vec{e} \times \vec{s}_{\mu}
\end{split}
\label{eq:U12}
\end{equation}
describes magnon modes with frequency $\omega_{\mu}$. Next, we calculate the magnon modes for the finite system using linear spin wave theory and obtain for wave vectors in the $z$-direction
\begin{equation}
\hbar \omega = \sqrt{ (2d_x - 2d_z + J_{\mathrm{eff}}) \cdot (2d_x + J_{\mathrm{eff}}) }
\label{eq:U13}
\end{equation}
with an effective exchange interaction term
\begin{equation}
J_{\mathrm{eff}} = D_{\mathrm{eff}} k^2
\label{eq:U14}
\end{equation}

Using the condition for the mode $n$ for standing waves (Eq. (\ref{eq:k})), we obtain the frequency in a bulk system as reference value.

\section{Results of the model}
\label{sec:modelresults}
In order to compare the analytical and numerical results (Appendix \ref{sec:appd}) to the experiments, we evaluate the magnon eigenmodes as function of the thickness of the magnetic layer. First, we discuss the magnon modes and the resulting effective exchange stiffness obtained by the effective one-dimensional model in thin films with vacancies randomly distributed around the boundary of the system. In the next step, we compare these results with results from atomistic spin dynamics simulations to verify the effective one-dimensional model and its underlying assumptions (Appendix \ref{sec:appd}). Finally, we explore the impact of modifications of the interfacial exchange parameters. We consider thicknesses between \SI{2}{\nano \metre} and \SI{10}{\nano \metre}.

\subsection{Thin films with vacancies at the boundary}

\begin{figure}[htbp]
	\centering
	\includegraphics[width=0.9\linewidth]{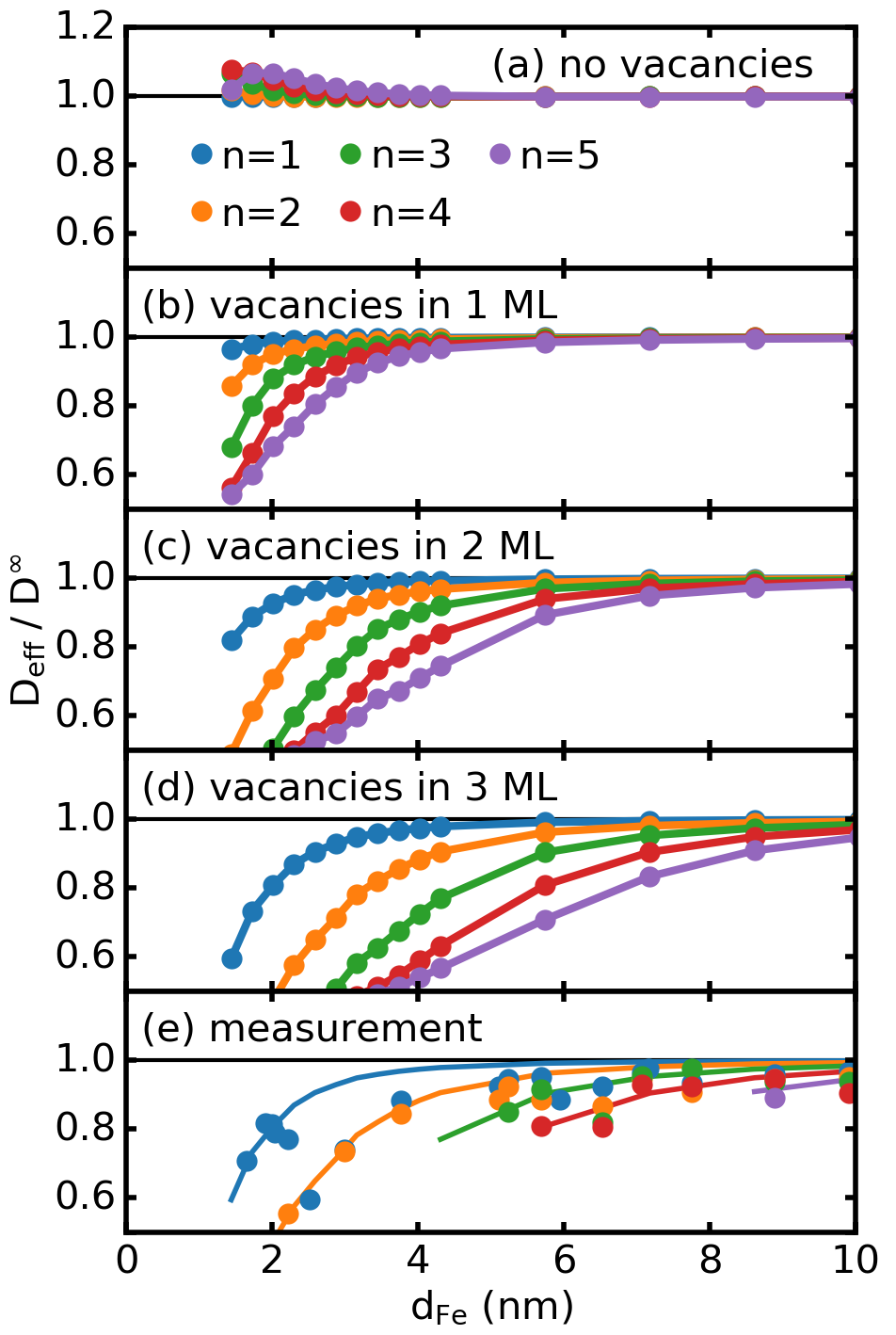}
	\caption{Effective exchange stiffness as a function of thickness
for a magnetic layer  without vacancies (a), for vacancies in 1 ML
(b), 2ML (c) and 3ML (d). Panel (e) shows the measured values as shown in Fig.~\ref{fig:stiff} normalized to $D^{\infty} =$ \SI{220}{\milli \eV \square \angstrom } (points) and for comparison analytical results for vacancies in 3 ML (lines).}
	\label{fig:U2}
\end{figure}
First, we consider the effective one-dimensional model and calculate the effective matrix $H_{zz'}$ from Eq. (\ref{eq:U9}) and determine the eigenvalues of the matrix for various thicknesses. Similar to the experimental evaluation, we calculate the effective exchange stiffness $D_{\mathrm{eff}}$ to describe the modes. The dispersion relation corresponding to the Hamiltonian can be approximated by
\begin{equation}
\hbar \omega = 2d_x + D_{\mathrm{eff}} k^2
\label{eq:U18}
\end{equation}
In all studied cases we consider that the wave vector $k$ of the $n$-th mode is given by the standing wave condition (Eq. (\ref{eq:k})). In Fig. \ref{fig:U2}, the resulting exchange stiffness is shown normalized to the theoretical bulk value. Fig.~\ref{fig:U2}(a) presents the effective exchange stiffness as a function of the thickness for several magnon modes for a homogeneous occupation of magnet lattice sites without vacancies. The exchange stiffness only shows a very small increase for thin films, due to the boundary conditions. Fig.~\ref{fig:U2}(b) shows the effective exchange stiffness for the case of vacancies in the first and last monolayer only. The reduction of the stiffness is much larger for higher magnon modes. The first magnon mode has almost constant stiffness for all thicknesses, whereas for the 4th magnon mode the stiffness is reduced by about 40\% for films with a thickness of \SI{2}{\nano \metre}. If we include more vacancies in the magnetic material, the thickness-dependent reduction further increases. This is demonstrated in Fig.~\ref{fig:U2} (c) and (d). Here, vacancies occur in the first two and three monolayers, respectively, around the layer boundaries. Again, the reduction of the effective exchange stiffness is more prominent for higher modes and the stiffness decreases by more than 60\%. 

The effect of stronger stiffness reduction for larger mode numbers is recognizable in the experimental dependencies too and the model curves from Fig.~\ref{fig:U2}(d) match the data quite well as highlighted by Fig.~\ref{fig:U2}(e). 
This vacancy-induced reduction of the exchange stiffness has its main origins in the matrix elements of $H_{zz'}$ which are lowered in the vicinity of the interfaces: the reduced number of neighboring Fe ions reduces the average exchange interaction there. 

Atomistic spin dynamics simulations (Appendix \ref{sec:appd}) show a similar reduction of the effective exchange stiffness for ultra-thin films. This demonstrates that the effective one-dimensional model, despite its simplifications, gives overall a good description of the magnon modes in thin magnetic layers. In the following, we will therefore consider only the effective one-dimensional model. 

\subsection{Impact of reduced exchange constants}

\begin{figure}[htbp]
	\centering
	\includegraphics[width=0.9\linewidth]{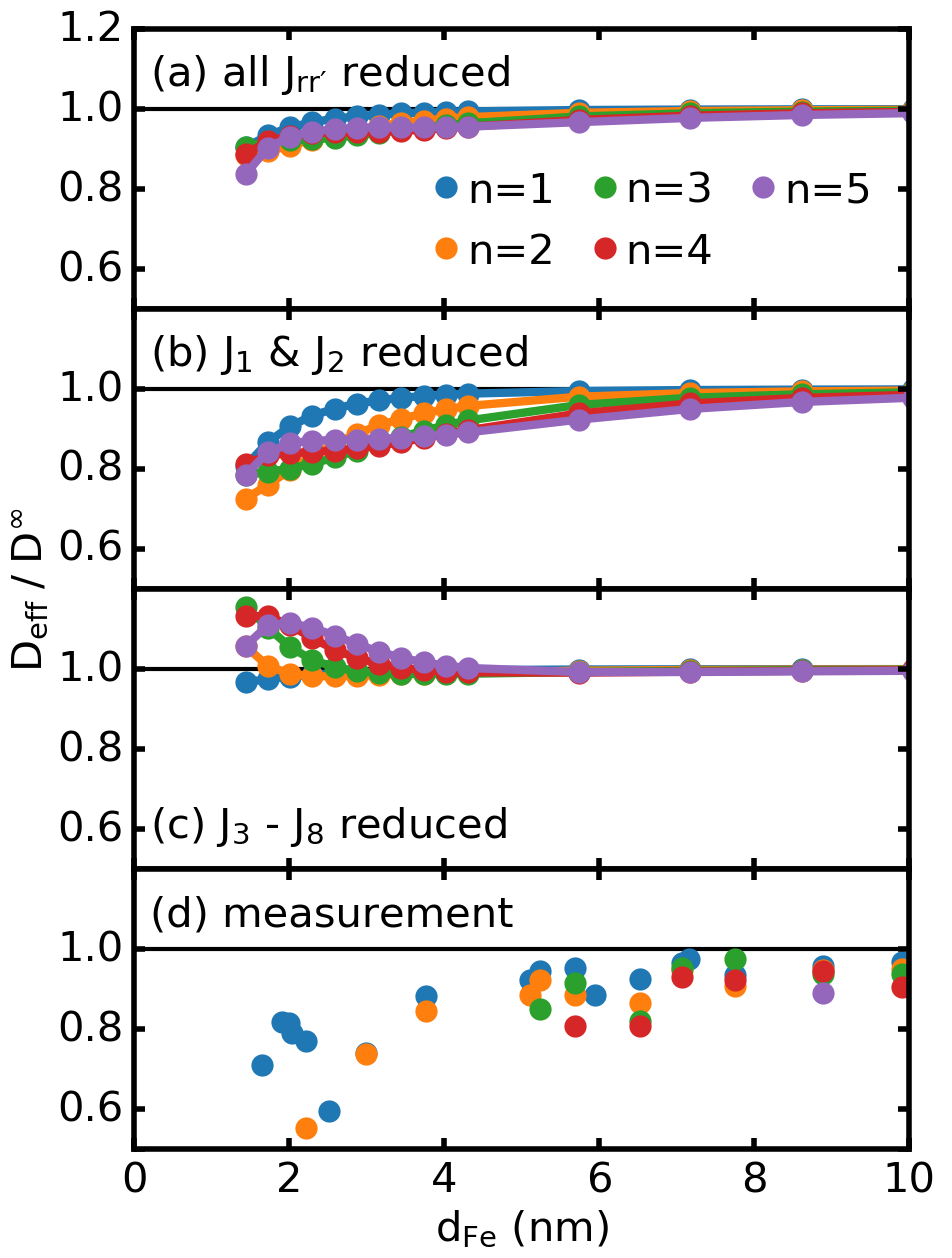}
	\caption{Effective exchange stiffness as a function of thickness for several magnon modes in magnetic materials where exchange constants are reduced at the interface. (a): All 8 parameters are reduced. (b): $J_1$ and $J_2$ are lowered. (c): $J_3$-$J_8$ are lowered. Note, that no changes of Fe concentration are considered here. (d): measured values as shown in Fig.~\ref{fig:stiff} normalized to $D^{\infty} = $ \SI{220}{\milli \eV \square \angstrom }}
	\label{fig:U5}
\end{figure}

At the interface the electronic band structure differs from the one in the bulk region, thus exchange parameters will be different as well. A calculation of these modifications is beyond the scope of this work, but we will demonstrate how modifications of the exchange parameters influence the magnon frequencies and the resulting effective exchange stiffness. The results are shown in Fig.~\ref{fig:U5}. We consider three different settings. In Fig.~\ref{fig:U5} (a), we show the results if all 8 exchange interaction parameters (see Table \ref{tab:input}) at the first and last layer are reduced to 60\% and the second and second last layer are reduced to 80\% (Fig.~\ref{fig:U1} (b)). Note that we reduce the parameter, if one of the interacting magnetic moments is within these layers and no vacancies are considered. For the thinnest films (2 nm), this leads to a stiffness reduction of about 10\%.

Next, we consider that the exchange coupling parameters will not all change equally at the interface, as they have different origins. To implement this, we only reduce the exchange interaction parameters $J_1$ and $J_2$, describing the nearest-neighbor interaction, at the interface similar to the first scenario. As illustrated in Fig.~\ref{fig:U5} (b), this leads to a significant reduction of about 30\%. The exchange interaction is strongest for nearest neighbors and the amplitude of the interaction decays if the two moments are further separated. The nearest neighbor exchange interaction has a larger amplitude than the effective exchange interaction including all interaction shells, therefore a reduction only within these layers has a larger effect than a reduction of all exchange interactions.
If we on the contrary reduce only the parameters $J_3$ up to $J_8$, we observe even an increase of the effective exchange. This is illustrated in Fig.~\ref{fig:U5} (c). Here, the exchange interaction parameters with the largest amplitudes are $J_4$ and $J_5$ and both are negative. Reducing the parameters at the boundaries therefore increases the effective exchange interaction. In all shown cases, the resulting reduction of the exchange stiffness occurs not only for the thinnest films but is also present for film thicknesses beyond 6~nm.

\section{Discussion}
\label{sec:discussion}
We measured the PSSWs by time-resolved MOKE at frequencies of up to \SI{2.4}{\tera \hertz} and up to the 5th mode as a function of thickness between 1 nm and 17 nm. From these measurements we obtain a bulk exchange stiffness constant for Fe that is about 20 \% lower compared to previously reported obtained by neutron scattering. Using an additional set of epitaxial iron layer samples with thicknesses between 30 nm and 85 nm we confirmed the reduced bulk value of the exchange stiffness for the PSSWs by FMR measurements. We would like to point out that the sample thicknesses were carefully verified using RHEED oscillations and RBS. Therefore, we are rather confident that the PSSWs in iron layers are indeed better described with an exchange stiffness that is 20\% lower than the literature value.

The concept of stiffness is introduced for sufficiently small wave vectors, where the magnon dispersion can be well approximated by a parabola near its minimum. The corresponding range of wave vectors is determined by the length of exchange interaction: it reduces with increasing the number of involved coordination spheres, which is particularly important in the case of Fe. At the same time, neutron scattering experiments typically have very large error bars at small energies and the exchange stiffness was determined in iron bulk crystals at magnon energies between 10~meV and 100~meV \cite{Shirane1968}. In this range even the used next-order correction ($\propto k^4$) may  not be sufficient: already at 10~meV its contribution exceeds 3\% and rises above 30\% at 100~meV if only the first two coordination spheres are considered. Moreover, at high energies the dispersion is additionally affected by hybridization between the spin wave band and the Stoner continuum \cite{Prokop2009}. On the contrary, our measurements are performed in a lower magnon energy range: from 10~meV down to \SI{160}{\micro \eV} (i.e. \SI{120}{\micro \eV} above the dispersion minimum corresponding to the FMR mode) in MOKE experiments and further down to less than \SI{5}{\micro \eV} above the minimum in the FMR experiments. The low-energy range (typically below 1~meV) is typically covered in BLS experiments (c.f. Ref.\cite{Hicken1995}) however the line broadening and the mixing of in- and out-of-plane wave vector components complicate the analysis. Therefore, the technique presented here (coherent excitation PSSWs with relatively large amplitude) complements the other methods in the range of small and moderate wave vectors. Currently, it offers the most accurate determination of the magnon stiffness because, on one hand, in this range the magnon dispersion is to a good extent quadratic (i.e. well-described by the stiffness), on the other hand, the range is large-enough to well determine the dispersion curvature (i.e. the stiffness).
 
Furthermore, for films with a thickness below $\approx$ \SI{10}{\nano \metre} an interface-induced reduction of the exchange stiffness is observed. This reduction is larger for higher PSSW modes. In order to understand this effect we studied various possible mechanisms which can lead to a modification of magnon frequencies in thin films: vacancies and spatially-dependent exchange interaction parameters. 
We conclude that interface alloying on the monolayer scale (modelled by vacancies) causes a strong reduction of the exchange stiffness, which is even stronger for higher PSSW modes in agreement with the experimental observation (Fig.~\ref{fig:U2}). A selective reduction of absolute values of exchange parameters can either increase or decrease the frequency of the magnon modes: Lowering the short range interaction leads to a reduction, whereas lowering the absolute value of long range exchange parameters can increase the magnon frequency due to their antiferromagnetic contributions (Fig.~\ref{fig:U5}). In both cases a mode splitting occurs, which is nearly compensated when all exchange parameters are reduced. The modification of the exchange stiffness due to vacancies is only significant for very thin films (few nanometer thick), whereas spatially dependent exchange constants can also show effects for thicker films. Although both effects can contribute, it is difficult to conclude on the role of modified exchange constants in the real system since our experimental data are already reasonably well described by the intermixing contribution alone. As shown in Fig.~\ref{fig:U2}~(e), the interdiffusion only up to 3~ML around the interface (consistent with the high epitaxial quality of the interfaces in our samples) explains the experiments. This result shows that careful modelling of the spinwave behavior can be used to estimate the alloying at interfaces.

Finally, the PSSW mode spectrum that is excited by the spin current pulses in our experiments allows us to estimate the depth profile of the spin transfer torque pulse. In a previous work \cite{Razdolski2017} we illustrated that the STT excitation depth is about a quarter of the shortest PSSW wavelength that can be excited:  $\lambda_{\mathrm{STT}} \lesssim 1/4 \, \lambda_{\mathrm{min}}$. Here our smallest wavelength occurs for the second spin-wave mode (2.4 THz) at $\lambda_{\mathrm{min}}=$2.22 nm, which means that STT excitation depth is $ \lambda_{\mathrm{STT}} \lesssim$ \SI{0.56}{\nano\metre} $\approx$ 4~ML. This result is consistent with the absorption length of the transverse spin component that one can expect for iron \cite{Stiles2002}.

\section{Conclusion}
\label{sec:conclusion}
Using a novel method to excite coherent spin waves in the THz frequency range we explored exchange stiffness. Our two main results are 
(i) the exchange stiffness for bulk iron relevant for spin waves with wavelengths between 2.2~nm and 100~nm actually has a value of  D = \SI{220+-10}{\milli \eV \square \angstrom }. This value is about 20 percent lower than the value extracted from neutron scattering.
(ii) we observed an interface-induced reduction of spin wave stiffness for very thin iron layers. By comparing the experimental results to a microscopic model we conclude that this effect can be explained by intermixing at interfaces on the monolayer scale. 

\begin{acknowledgements}
Financial support from the German research foundation (DFG) through collaborative research center CRC/TRR 227 (Project B01) is greatfully acknowledged.
IR acknowledges the National Science Centre Poland (Grant No. 2019/35/B/ST3/00853)
\end{acknowledgements}

\appendix

\section{Sample and experimental setup}
\label{sec:appa}
The Au-layer of the sample has the thickness of \SI{70}{\nano\metre} at one half of the sample (measurement area) and only \SI{5}{\nano\metre} at other half (alignment area).  The alignment area is required to adjust and maintain the spatial and temporal overlap of pump and probe laser pulses.
 
Pump and probe beams were modulated (chopped) with different frequencies and focused using off-axis parabolic mirrors into a $\approx$ \SI{20}{\micro\metre} spot, which results in a pump fluence of $\approx$ \SI{5}{\milli\joule\per\square\centi\metre}.  The reflected second harmonic (SH) signal was separated by a dichroic beam splitter, passed a monochromator, and registered by a photomultiplier operating in the photon counting regime. A small fraction of the fundamental reflected light was  directed to a photodiode to measure the transient linear reflectivity (LR). The remaining reflected light was split in a 1:1 ratio in two branches where MOKE rotation and ellipticity signals were measured simultaneously. Here a balanced detection scheme was realized with the help of a Glan-laser prism and two photodiodes. In the ellipticity branch, a quarter-wave plate was installed before the prism.

Two crossed translators allowed the precise  motion of the sample exactly parallel to its plane. The edge of the wedge and the border between measurement and alignment areas were identified with both LR and SH signals.

For each measurement, the collector thickness was set by displacing the sample horizontally to the corresponding distance between the laser spot and the edge of the wedge. At the alignment area, the spatial overlap of pump and probe beams was optimized by maximizing the transient LR response at positive delays and the zero time delay was determined precisely by maximizing the SH cross-correlation signal. After that the sample was translated vertically to bring the spot to the measurement area (here only a small translation is required) and hysteresis loops in SH signals from reflected probe (detecting the magnetization direction in the collector) and reflected pump (detecting the magnetization direction in the emitter) beams were recorded simultaneously (for different phases of light choppers) by scanning the vertical (transverse) magnetic field at various values of horizontal (longitudinal) field \cite{Alekhin2017}.

\section{Magnetic characterization}
\label{sec:appb}
\begin{figure}[htbp]
		\includegraphics[width=0.9\linewidth]{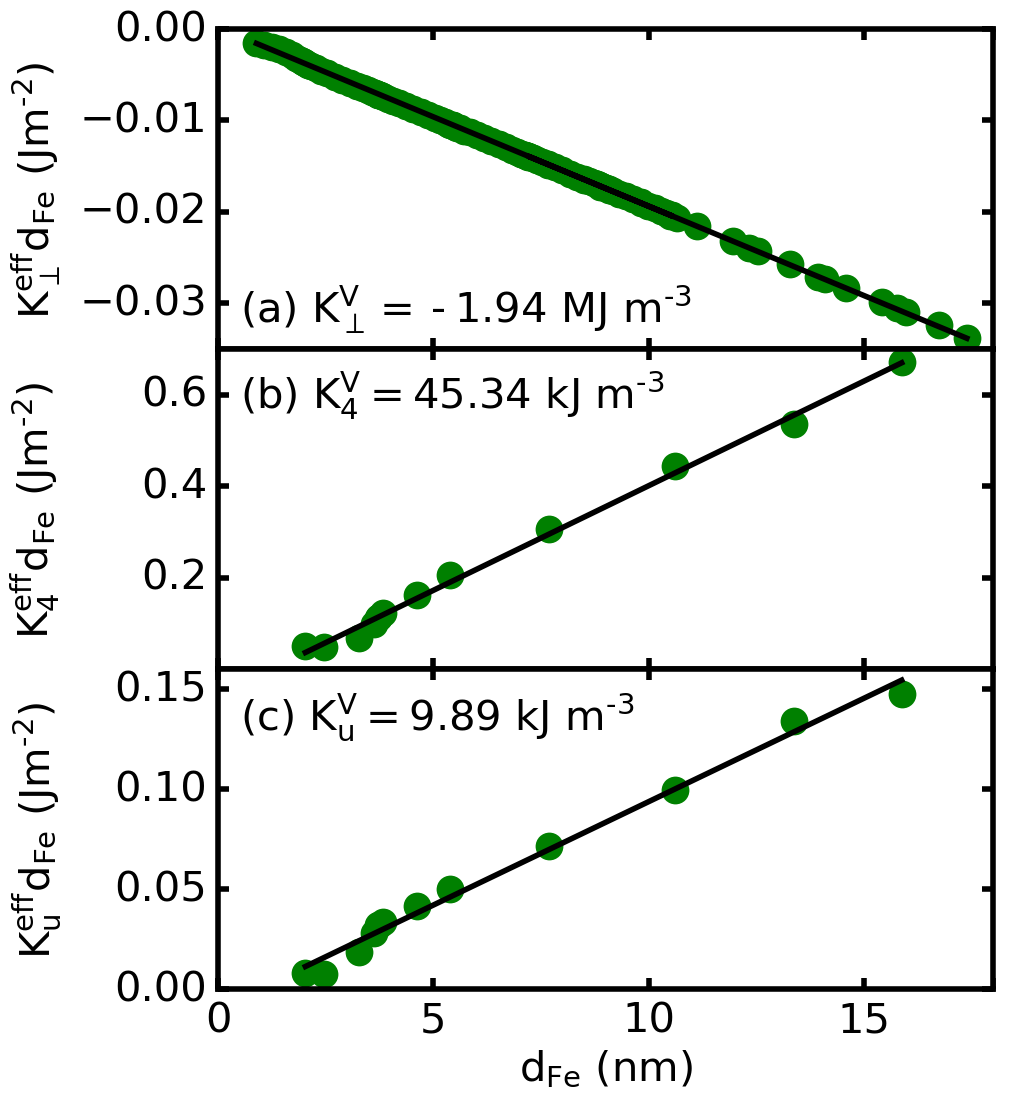}
	\caption{(a) Effective out-of-plane anisotropy constant $K_{\perp}^{\mathrm{eff}} d_{\mathrm{Fe}}$ obtained by out-of-plane static MOKE measurements.
	(b) effective fourfold anisotropy constant $K_{4}^{\mathrm{eff}} d_{\mathrm{Fe}}$ and (c) uniaxial anisotropy constant $K_{u}^{\mathrm{eff}} d_{\mathrm{Fe}}$ obtained by in-plane static MOKE measurements}
	\label{fig:staticMOKE}
\end{figure}

For the analysis above it is necessary to know the anisotropy and demagnetization fields as a function of the thickness $d_{\mathrm{Fe}}$ of the (ultrathin) Fe-collector layer. From out-of-plane and in-plane static MOKE measurements values for the out-of-plane $K_{\perp}^{\mathrm{eff}}$ as well as fourfold $K_{4}^{\mathrm{eff}}$ and uniaxial (in-plane) $K_{\mathrm{u}}^{\mathrm{eff}}$ anisotropy constants are obtained. The bulk $K_V$ and interface $K_I$ contributions are separated by fitting the linear thickness dependence of the effective anisotropy: $K^{\mathrm{eff}}_{\perp,4,u}d_{\mathrm{Fe}}=K^V_{\perp,4,u} d_{\mathrm{Fe}} + K^I_{\perp,4,u}$ (Fig.~\ref{fig:staticMOKE}). 

From this we obtain a bulk-demagnetization field of 2.2 T and a bulk value of anisotropy field which consists of both fourfold and uniaxial anisotropy of 65 mT. It is well known that the interface magneto-crystalline anisotropy favours the perpendicular (i.e. out-of-plane) direction \cite{Yang2011,Hallal2013}, therefore $|K^{\mathrm{eff}}|$ decreases with decreasing iron thickness. A similar behavior is observed for the in-plane magneto-crystalline anisotropy. Their decrease for thinner samples originates in a volume  and an interface  terms with two competing easy axes oriented along [001] and [110], respectively \cite{Heinrich1991}. Both, the critical thicknesses in Eq. (\ref{eq:Handem}) corresponding to $K^{\mathrm{eff}}=0$ for anisotropy and demagnetization fields yield to $d_{\mathrm{an}}^0 \approx d_{\mathrm{dem}}^0 \approx$ \SI{1}{\nano \metre}. This critical thickness coincides with literature values of spin reorientation transition in iron observed for similar layer systems \cite{Brockmann1997}.

\section{FMR results with thicker Fe layers}
\label{sec:appc}
\begin{figure}[htbp]
\centering
		\includegraphics[width=0.9\linewidth]{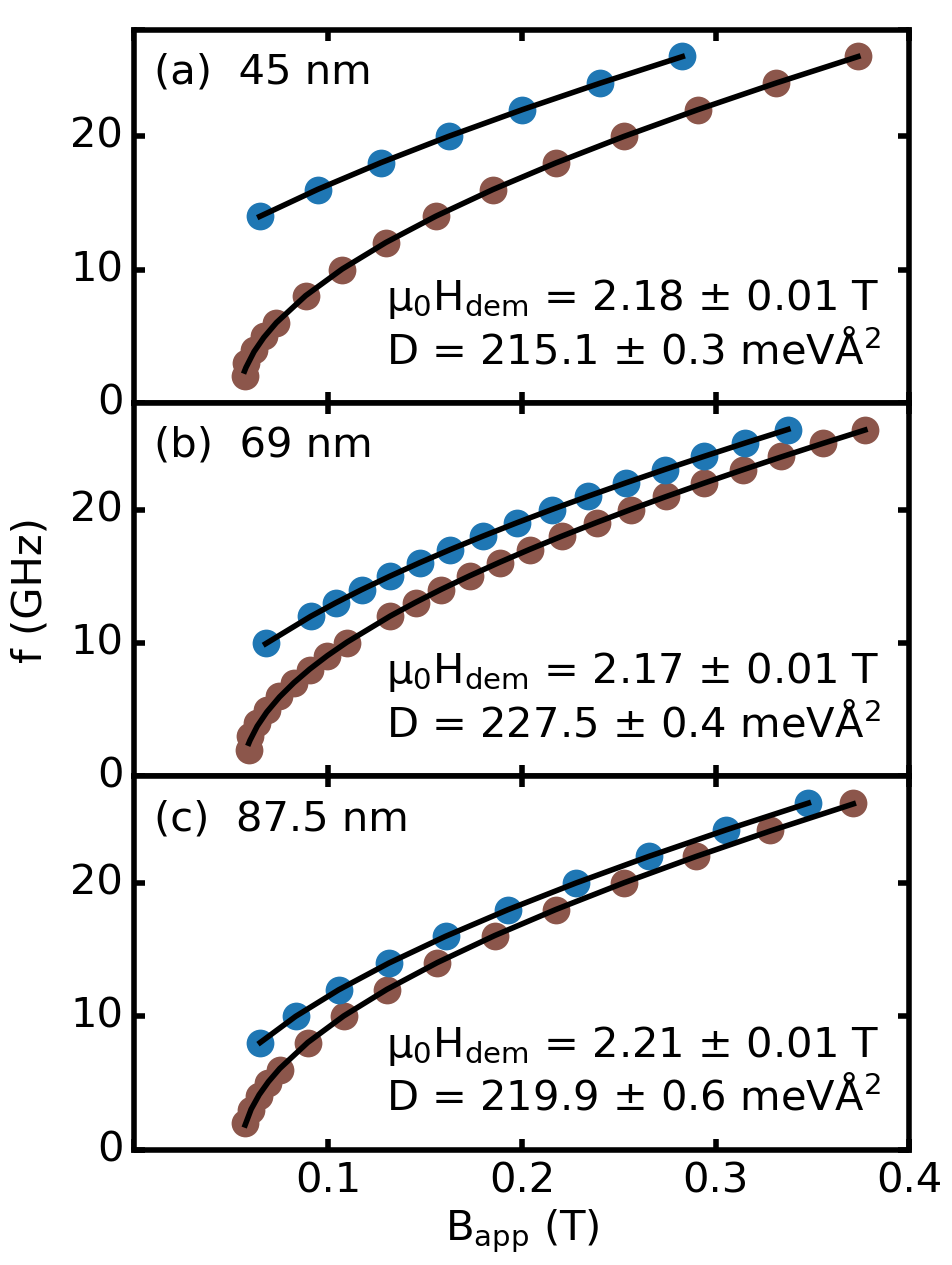}
	\caption{Measured resonance fields for FMR (brown points) and first PSSW (blue points) modes and resulting fit (black lines) for \SI{45}{\nano \metre} (a), \SI{69}{\nano \metre} (b) and \SI{87.5}{\nano \metre} (c) thick iron samples. The error bars for thickness and demagnetization field are only the result of the fit.}
	\label{fig:FMR}
\end{figure}

For the samples with iron thickness 45 nm, 69 nm, and 87.5 nm the  FMR mode and the first PSSW modes were resonantly excited via a coplanar waveguide in the range between 2 - 26 GHz in an external in-plane magnetic field $B_{\mathrm{app}}$. FMR spectra  were recorded using field  modulation and lock-in detection. The measured resonance fields at given frequencies for FMR and first PSSW modes were fitted using Eq. (\ref{eq:spinwave}) including an additional term accounting for the applied in-plane magnetic field $B_{\mathrm{app}}$. Resulting demagnetization fields and stiffness constants for the different samples are shown in the corresponding panels of Fig.~\ref{fig:FMR}(a)-(c).
The error bars for the spin wave stiffness from the FMR measurements in Fig.~\ref{fig:stiff} were estimated to $\pm$~\SI{8}{\milli \eV \square \angstrom} and originate from uncertainties of the sample thickness $\Delta d_{\mathrm{Fe}}=$ \SI{+-0.5}{\nano \metre} (based on RBS results), anisotropy fields $\Delta H_{\mathrm{an}}=$ \SI{+-1}{\milli \tesla} and demagnetization fields $\Delta H_{\mathrm{dem}}=$ \SI{+-0.1}{\tesla}.

\section{Atomistic spin dynamics simulations}
\label{sec:appd}
As a complementary approach, we perform numerical simulations of a magnetic thin layer excited by ultrafast spin-transfer torques. The configuration is similar to previous theoretical studies of laser excited standing waves in trilayer systems \cite{Ritzmann2020,Ulrichs2018}. We consider a magnetic thin layer with variable thickness $d$ and a cross-section of \SI{8.61}{\nano \metre} $\times$ \SI{8.61}{\nano \metre} with periodic boundary condition and integrate numerically the equation of motion including a phenomenological damping term $\alpha= 0.001$ and an ultrafast spin-transfer torque term acting only on the 5th layer by using atomistic spin dynamics simulations with a Heun solver. Since the Fe concentration at the boundary is reduced, we implement the excitation close to the boundary, but at a layer with fully occupied lattice sites. The equation of motion is
\begin{equation}
\begin{split}
\frac{\partial \vec{S}_i}{\partial t} = 
&- \frac{1}{\hbar} \vec{S}_i (t) \times \vec{H}_i (t) + \alpha \vec{S}_i (t) \times \frac{\partial \vec{S}_i (t)}{\partial t} \\
&+ j_{\mathrm{IF}} (t) \vec{S}_i (t) \times [\vec{S}_i (t) \times \hat{\vec{z}}]
%&- j_{\mathrm{IF}} (t) \vec{S}_i (t) \times [\vec{S}_i (t) \times \hat{\vec{m}}]
\end{split}
\label{eq:U16}
\end{equation}
where $j_{\mathrm{IF}}$ is the absolute value of spin current transferred at the interface and is given by
\begin{equation}
j_{\mathrm{IF}} = j_0 \cdot \frac{\exp{(-t/\tau_2)}}{1+\exp{(-(t-t_0)/\tau_1)}} \delta(z-\frac{5a}{2})
\label{eq:U17}
\end{equation}
similar as in a previous work \cite{Ulrichs2018}. Consistent with the experiments \cite{Razdolski2017} we use $j_0 = 0.1$, $t_0 =$ \SI{50}{\femto \second}, $\tau_1 =$ \SI{10}{\femto \second} and $\tau_2 =$
\SI{150}{\femto \second}. We solve the equation of motion for each magnetic moment and average the resulting magnetization over the cross-section for a fixed position $z$.

\begin{figure}[htbp]
\centering
		\includegraphics[width=1\linewidth]{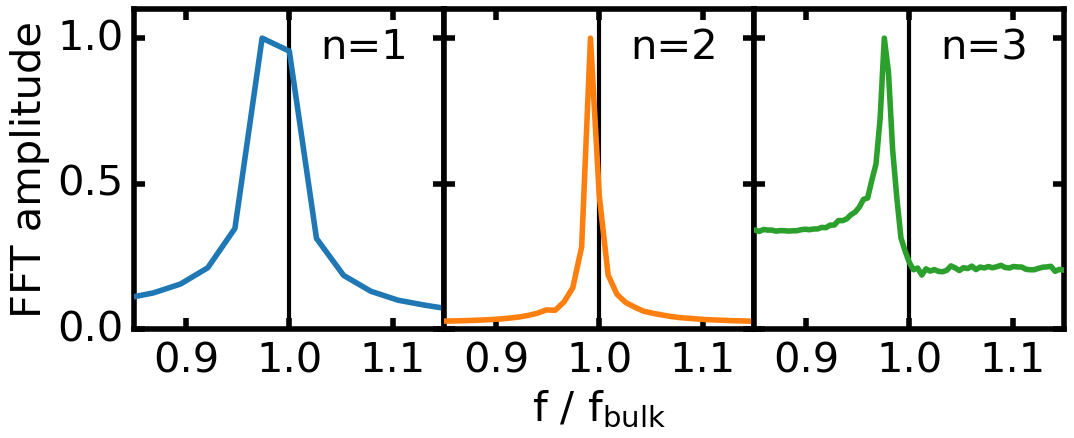}
	\caption{Resulting frequency spectrum for 30 ML (\SI{4.3}{\nano \metre}) iron with vacancies in 1 ML obtained by atomistic spin dynamics simulation. Amplitudes are normalized to peak amplitude and frequencies are normalized to bulk values.}
	\label{fig:simulation}
\end{figure}

We study the temporal evolution within the magnetic thin film after excitation with ultrafast spin-transfer torque over a time range of \SI{20}{\pico \second}. In Fig. \ref{fig:simulation} we compare the homogeneous film to spatially dependent concentration profiles as shown in Fig.~\ref{fig:U1}(a). We calculate the average magnetization for a fixed position $z$ and perform a Fourier transformation in the time domain to extract the spin wave excitations in the system. Since the concentration profile is realized by randomly placed vacancies we average over 10 different configurations. The results summarized in Fig.~\ref{fig:simulation} show a small reduction of the frequency for thin films in agreement with the analytical model. Note that peaks of the higher modes are not resolved for the thinnest films, due to a fast decay of these modes in the simulations. Nonetheless, the simulations show a reduction of the effective exchange stiffness for ultra-thin films. This reduction is considerably stronger for higher modes. This demonstrates that the effective one-dimensional model, despite its simplifications, gives overall a good description of the magnon modes in thin magnetic layers.

%\newpage
%\clearpage
%apsrev4-2.bst 2019-01-14 (MD) hand-edited version of apsrev4-1.bst
%Control: key (0)
%Control: author (72) initials jnrlst
%Control: editor formatted (1) identically to author
%Control: production of article title (-1) disabled
%Control: page (0) single
%Control: year (1) truncated
%Control: production of eprint (0) enabled
%

\end{document}